\documentclass{llncs}

\usepackage[dvips]{epsfig}

\begin{document}

\def\bottomfraction{.5}

\title{Investigation of the dynamical structure factor of the
Nagel-Schreckenberg traffic flow model}

\author{S.~L\"ubeck \and L.~Roters \and {K.\,D.}~Usadel}

%\author{S.~L\"ubeck\inst{1} \and M.~Schreckenberg\inst{1} 
%\and {K.\,D.}~Usadel\inst{1}}

\institute{Theoretische Physik, Gerhard-Mercator-Universit\"at, 
47048 Duisburg, Deutschland
%\and
%Universit\'{e} de Paris-Sud,
%Laboratoire d'Analyse Num\'{e}rique, B\^{a}timent 425,\\
%F-91405 Orsay Cedex, France
}

\maketitle

\markboth{\rm \tiny{
in {\it Traffic and Granular Flow 97},
edited by D.~E.~Wolf and M.~Schreckenberg,
Springer, Singapore (1998)}}
{\rm \tiny{
in {\it Traffic and Granular Flow 97},
edited by D.~E.~Wolf and M.~Schreckenberg,
Springer, Singapore (1998)}}
\thispagestyle{myheadings}
\pagestyle{myheadings}

\begin{abstract}
The Nagel-Schreckenberg traffic flow model
shows a transition from a free flow regime
to a jammed regime for increasing car density.
The measurement of the dynamical structure factor
offers the chance to observe the evolution of jams 
without the necessity to define a car to be jammed
or not.
Above the jamming transition the dynamical structure
factor $S(k,\omega)$ exhibits for a given $k$-value two maxima 
corresponding to the separation of the system into the 
free flow phase and jammed phase.
Analyzing the $k$-dependence of these maxima the backward 
velocity of the jams is measured.
We find that the jam  velocity neither depends on 
the global density of the cars nor the maximal velocity
of the model.
\end{abstract}

\section{Introduction}

In 1992 Nagel and Schreckenberg~\cite{NASCH_1} introduced a simple cellular
automata model, which simulates single-lane one-way traffic,
in order to study the transition from free flow traffic
to jammed traffic with increasing car density.
The behavior of the model is determined
by three parameters, the maximal velocity $v_{\rm max}$,
the noise parameter $P$ and the global density of cars $\rho=N/L$,
where $N$ denotes the total number of the cars and $L$ the
system size, respectively.
The variables describing a car $i$ at time $t$ are its position
$r_i \in\{ 1,2,...,L \} $, its velocity
$v_i \in \{0,1,...,v_{\rm max} \} $
and the gap $g_i$, which is 
the number of unoccupied cells in front of the car. 
Using periodic boundary conditions, the following update steps 
are applied in parallel for each car: 
\begin{eqnarray}
  v_i\;     & \longmapsto & \; \min(v_{\rm max}, v_i+1),    \nonumber\\
  v_i\;     & \longmapsto & \; \min(v_i, g_i),         \nonumber\\
  v_i\;     & \longmapsto & \; \max(0, v_i-1) \mbox{ \quad with probability }P,  \nonumber\\   
  r_i\;     & \longmapsto & \; r_i+v_i\,.     \nonumber
\label{eq:def_update_rules}
\end{eqnarray}

Increasing the global density for fixed $P$ and $v_{\rm max}$ 
jams occur above a certain critical value $\rho_c$~\cite{NASCH_1}. 
Contrary to the forward movement of all particles
the jammed region is characterized by a backward moving
of shock waves, i.e.~jams are nothing else
than backward moving density fluctuations.
This property of jams was found in real
traffic flow already in the 50's~\cite{LIGHT_1}.
Investigations of the Nagel-Schreckenberg
traffic flow model show that crossing the
critical point a transition takes place from a 
homogeneous regime (free flow phase) to an inhomogeneous 
regime which is characterized by a coexistence of
two phases (free flow and jammed traffic).
Thereby, the free flow is characterized by a low local
density and the jammed phase by a high
local density, respectively.
Due to the particle conservation of the model
the transition is realized by the system separating
into a low density region 
and a high density region~\cite{LUEB_6}.
Therefore, the dynamical structure factor is an appropriate 
tool to investigate the decomposition of the
two phases above the transition point.
Additionally the dynamical structure factor has
the decisive advantage that both phases
could be distinguished in a natural way 
by the sign of their characteristic velocities.
Thus no artificial definitions of jams are needed
which were used in previous investigations
and lead to controversial
results~(see for instance \cite{LUEB_6,CSANYI_1,SASVARI_1,EISEN_2}).

\section{Simulation and Results}

\begin{figure}[b]
 \begin{center}
 \epsfxsize=6.0cm
 \epsfysize=6.0cm
% \epsffile{/nfs/nujunkum/sven/traffic/ps_files/space_time_dia_02.eps}
 \epsffile{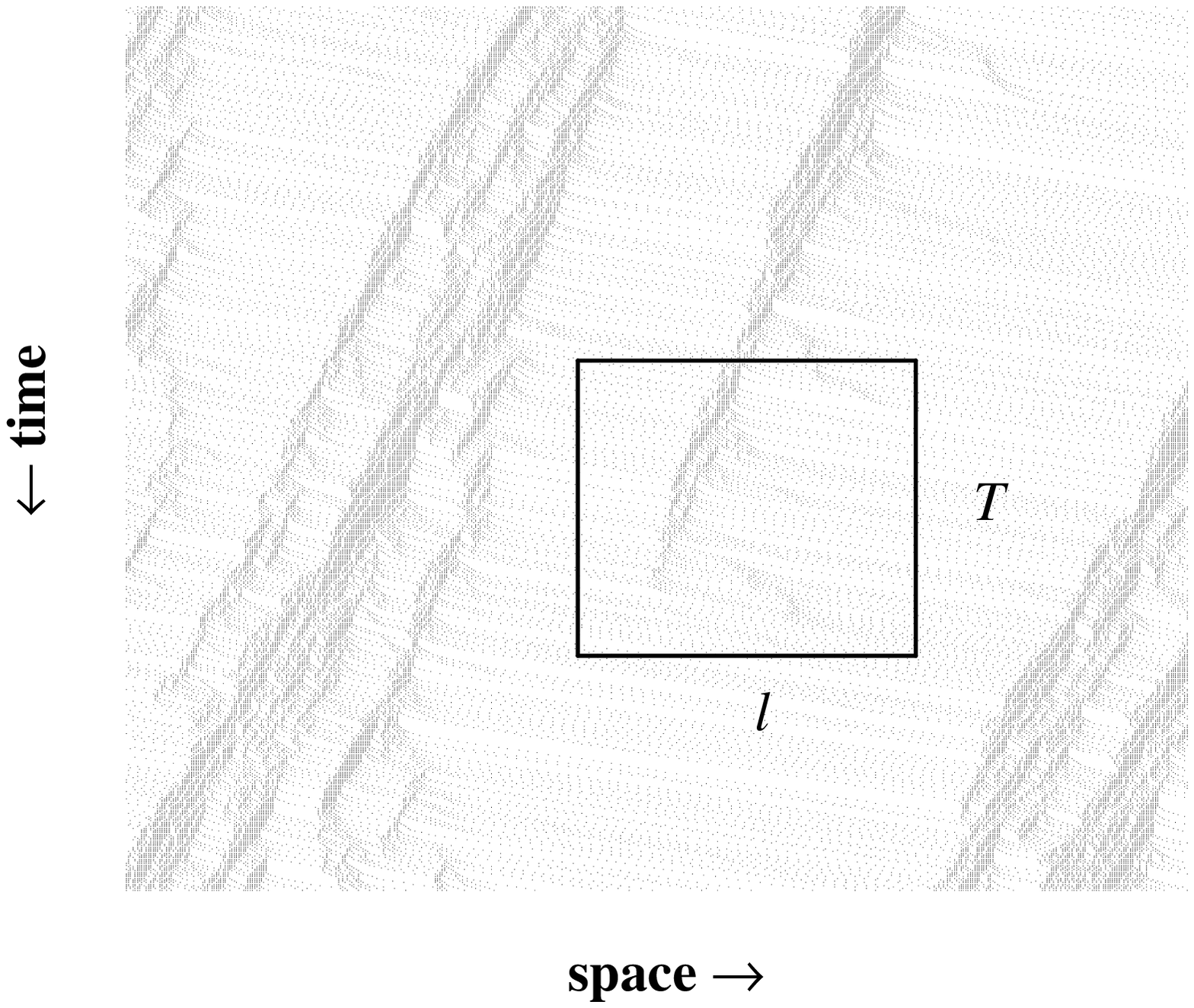}
 \end{center}
 \vspace{-0.8cm}
 \caption{Space-time plot for $v_{\rm max}=5$, $P=0.5$, and $\rho>\rho_c$.
          Note the separation of the system in high and low density
	  regions.}
 \label{fig:space_time} 
\end{figure}

The dynamical structure factor is defined as follows:
Let
\begin{equation}
\eta_{r,t} \; = \; \left \{ 
\begin{array}{l}
1 \hspace{0.5cm} \mbox{if cell $r$ is occupied at time $t$}\\
0 \hspace{0.5cm} \mbox{otherwise} \, .
\end{array} \right.
\end{equation}
The evolution of $\eta_{r,t}$ leads directly to the
space-time diagram where the propagation of the 
particles can be visualized.
Figure~\ref{fig:space_time} shows a space-time
diagram of the system above the critical value.
Traffic jams are characterized by the 
backward movement of the high density regions.
The dynamical structure factor $S(k, \omega)$ is given by
\begin{equation}
S(k,\omega) \; = \; l\,T\;\left \langle 
\left | \frac{1}{l T} \;{\displaystyle\sum_{r,t}} \, \eta_{r,t} \; 
e^{i (k r -\omega t) } \right |^2
\right\rangle \, ,
\label{eq:def_steady_state}
\end{equation}
where the Fourier transform is taken over a finite
rectangle of the space-time diagram of size $l\times T$ 
(see Fig.~\ref{fig:space_time}).

\begin{figure}[t]
 \begin{center}
 \epsfxsize=10.0cm
 \epsfysize=6.0cm
% \epsffile{/nfs/nujunkum/sven/traffic/paper_2/l64r_Vm5_rg00397_P0500.ps} 
 \epsffile{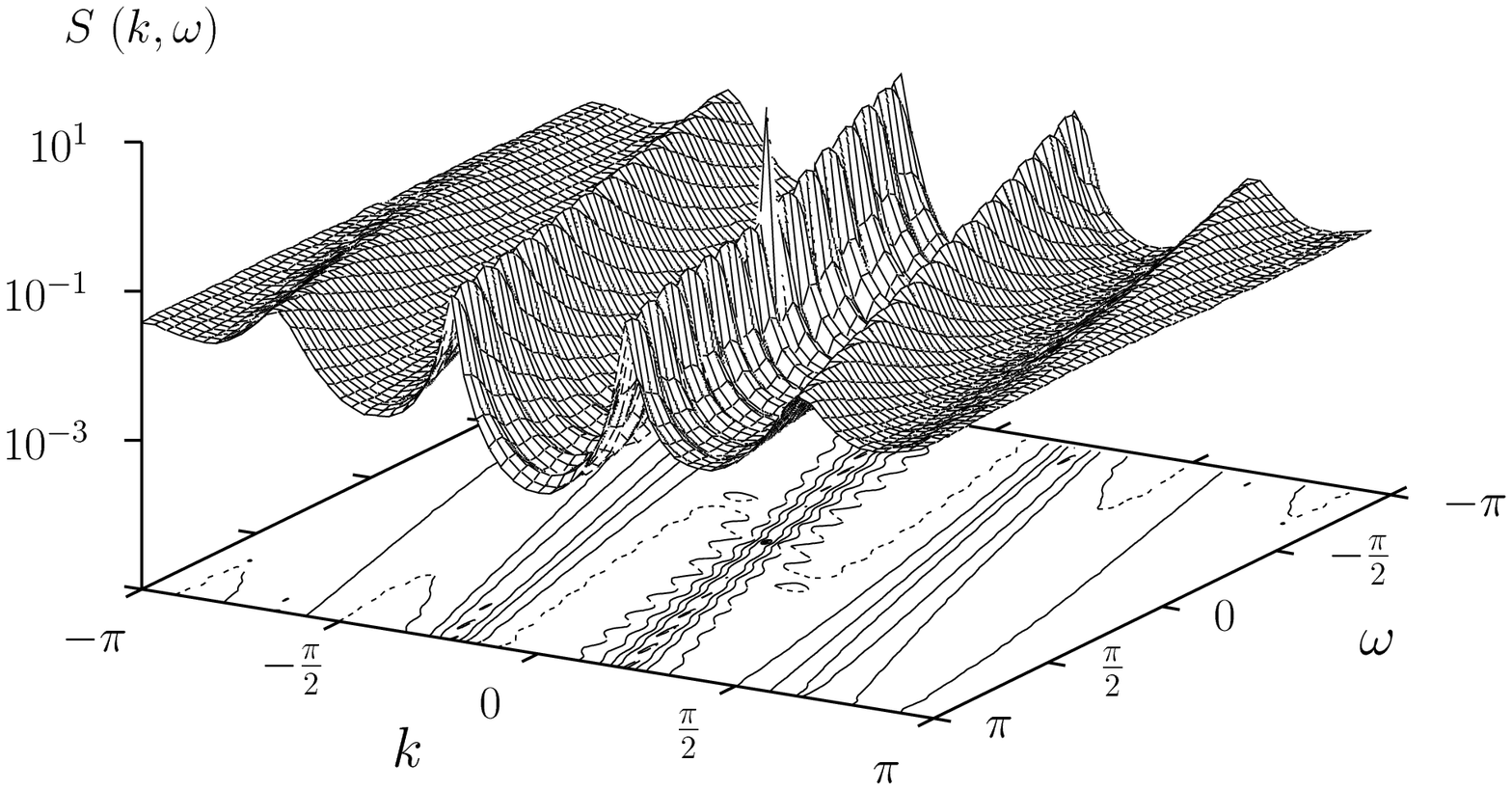} 
 \end{center}
 \vspace{-0.8cm}
 \begin{center}
 \epsfxsize=10.0cm
 \epsfysize=6.0cm
% \epsffile{/nfs/nujunkum/sven/traffic/paper_2/l64r_Vm5_rg00998_P0500.ps} 
 \epsffile{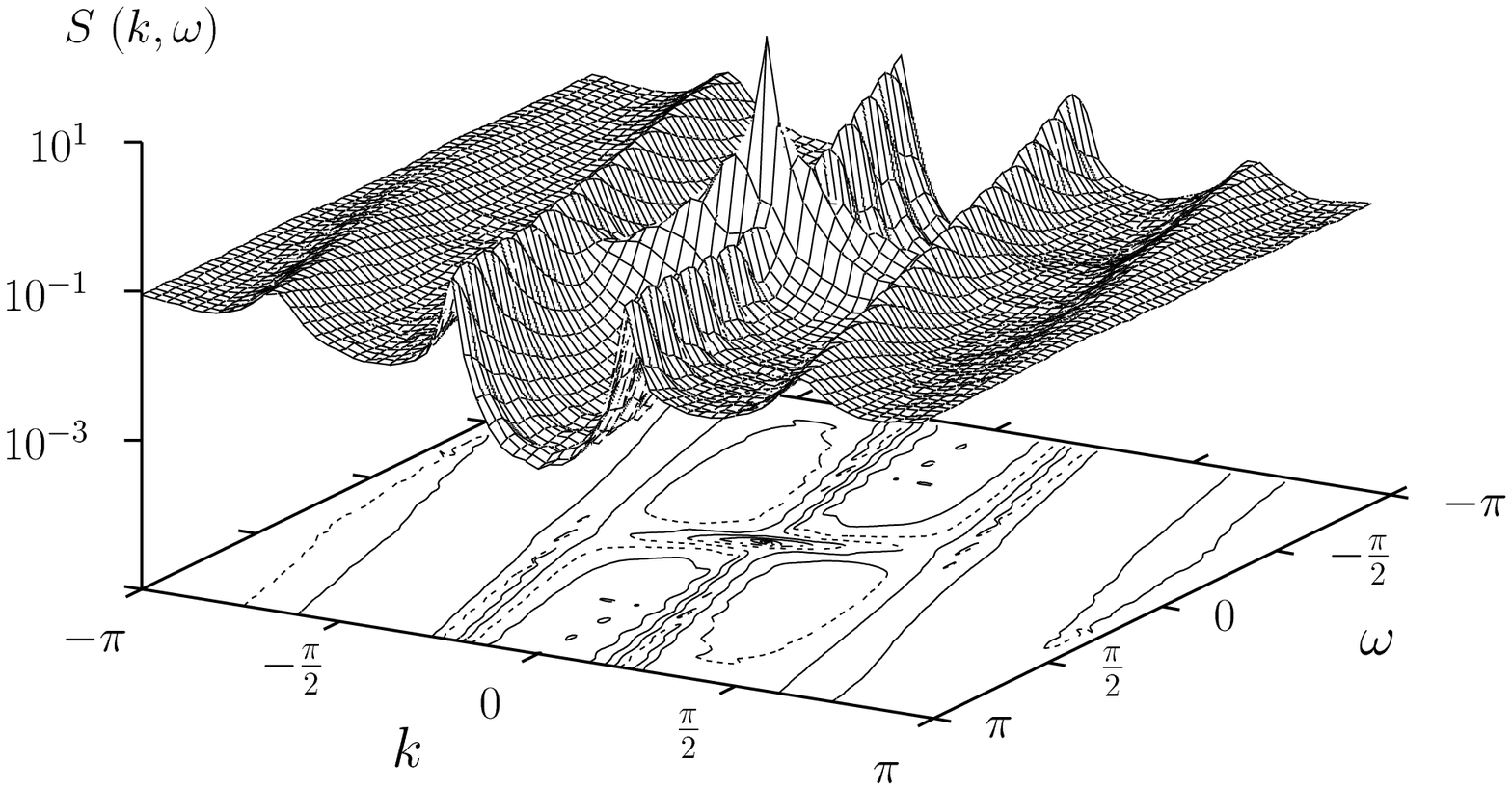} 
 \end{center}
 \vspace{-0.8cm}
 \caption{The dynamical structure factor $S(k,\omega)$ below (upper figure)
	  and above (lower figure) the critical value.
          The ridges with maximal $S(k,\omega)$ 
          indicates the various modes.}
 \label{fig:dyn_struc_factor} 
\end{figure}

In Fig.~\ref{fig:dyn_struc_factor} we plot the dynamical
structure factor below and above the transition, respectively.
Below the transition $S(k, \omega)$ exhibits one mode
formed by the ridges.
The values of the modes are obtained from a
determination of the position of the maxima of $S(k,\omega)$
for fixed $\omega$ or $k$, respectively.
This mode is characterized by a positive slope 
$v=\frac{\partial \omega}{\partial k}$ corresponding 
to the positive velocity of the particles
in the free flow phase. 
Increasing the global density of the cars, a second mode
appears at the transition to the coexistence regime. 
This second mode exhibits a negative slope
indicating that it corresponds to the backward moving 
density fluctuations in the jammed phase.

\begin{figure}[t]
 \begin{center}
 \epsfxsize=5.5cm
 \epsfysize=5.5cm
% \epsffile{/nfs/nujunkum/sven/traffic/ps_files/S_dyn_r_04.eps} 
 \epsffile{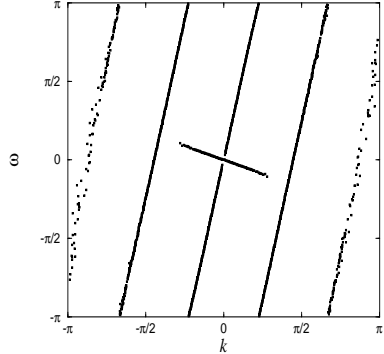} 
 \end{center}
 \vspace{-0.8cm}
 \caption{The maximal values of the dynamical structure
	  factor for $v_{\rm max}=5$, $P=0.5$, and $\rho>\rho_c$.
	  The lines with positive slope corresponds to the free flow
	  phase and the negative slope corresponds to the 
	  jammed phase, respectively.}
 \label{fig:dyn_struc_factor_max} 
\end{figure}

In Fig.~\ref{fig:dyn_struc_factor_max} we plot the position
of the ridges of $S(k, \omega)$ in the
jammed regime.
Both modes are characterized by a linear dispersion 
relation ($\frac{\partial \omega}{\partial k}={\rm const}$),
i.e.~each phase is characterized by one velocity only.
Since the modes of both phases can clearly be distinguished
(note the logarithmic scale in Fig.~\ref{fig:dyn_struc_factor})
and no other mode occur we think that this indicates 
that above the
transition a real phase separation of the system takes place.
This is in contrast to results of previous investigations
which examine the steady state correlation
function and concluded that there is no real
phase separation of the system~\cite{EISEN_2}.

\begin{figure}[b]
{\parbox{12cm}{
\flushbottom
\begin{minipage}{5cm}
 \epsfxsize=5.0cm
 \epsfysize=5.0cm 
\hspace{0.5cm}
% \epsffile{/nfs/nujunkum/sven/traffic/ps_files/v_free_02.eps} 
 \epsffile{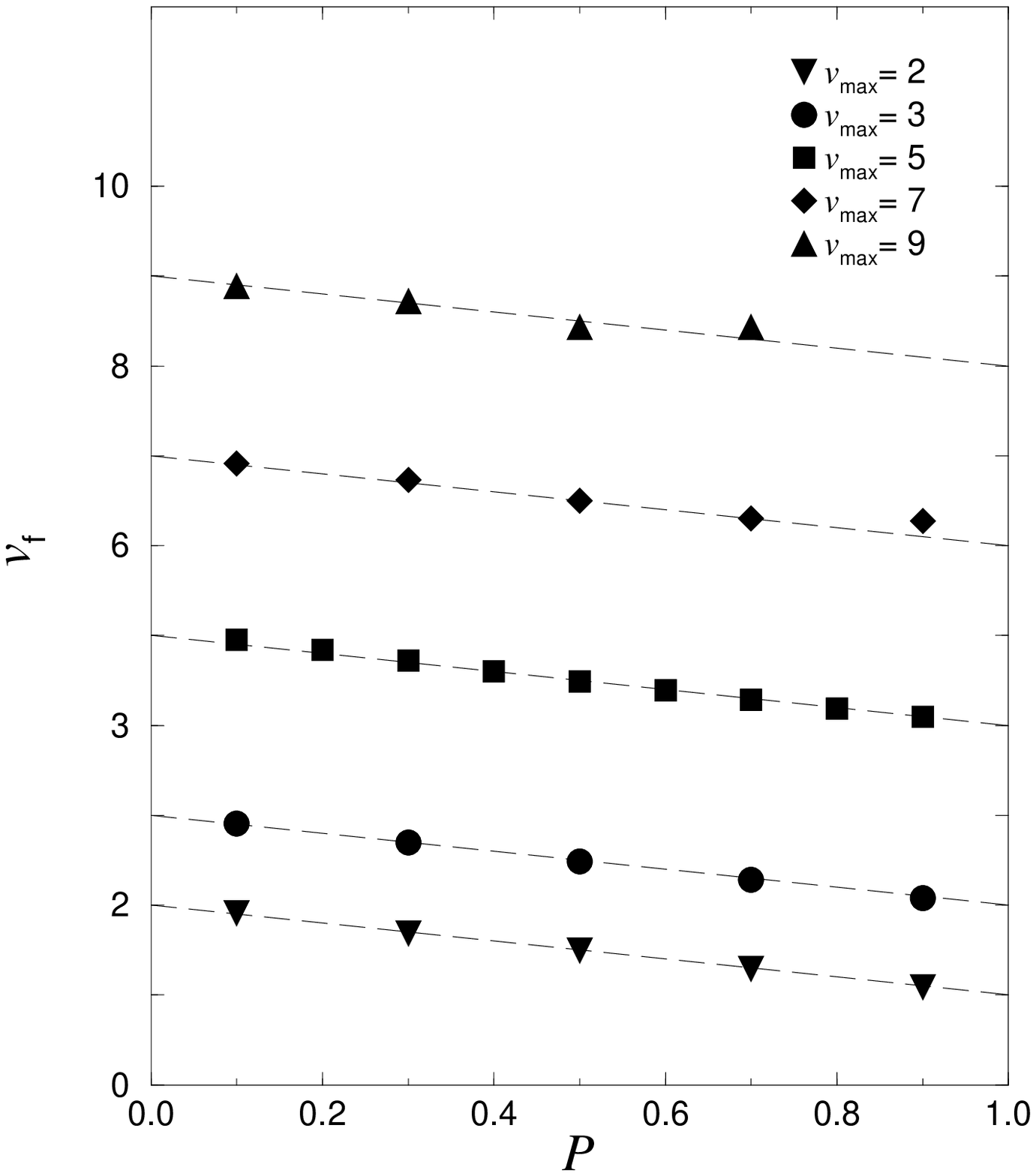} 
\end{minipage}
\hspace{0.5cm}
\begin{minipage}{5cm}
 \epsfxsize=5.0cm
 \epsfysize=5.0cm
% \epsffile{/nfs/nujunkum/sven/traffic/ps_files/v_jam_02.eps} 
 \epsffile{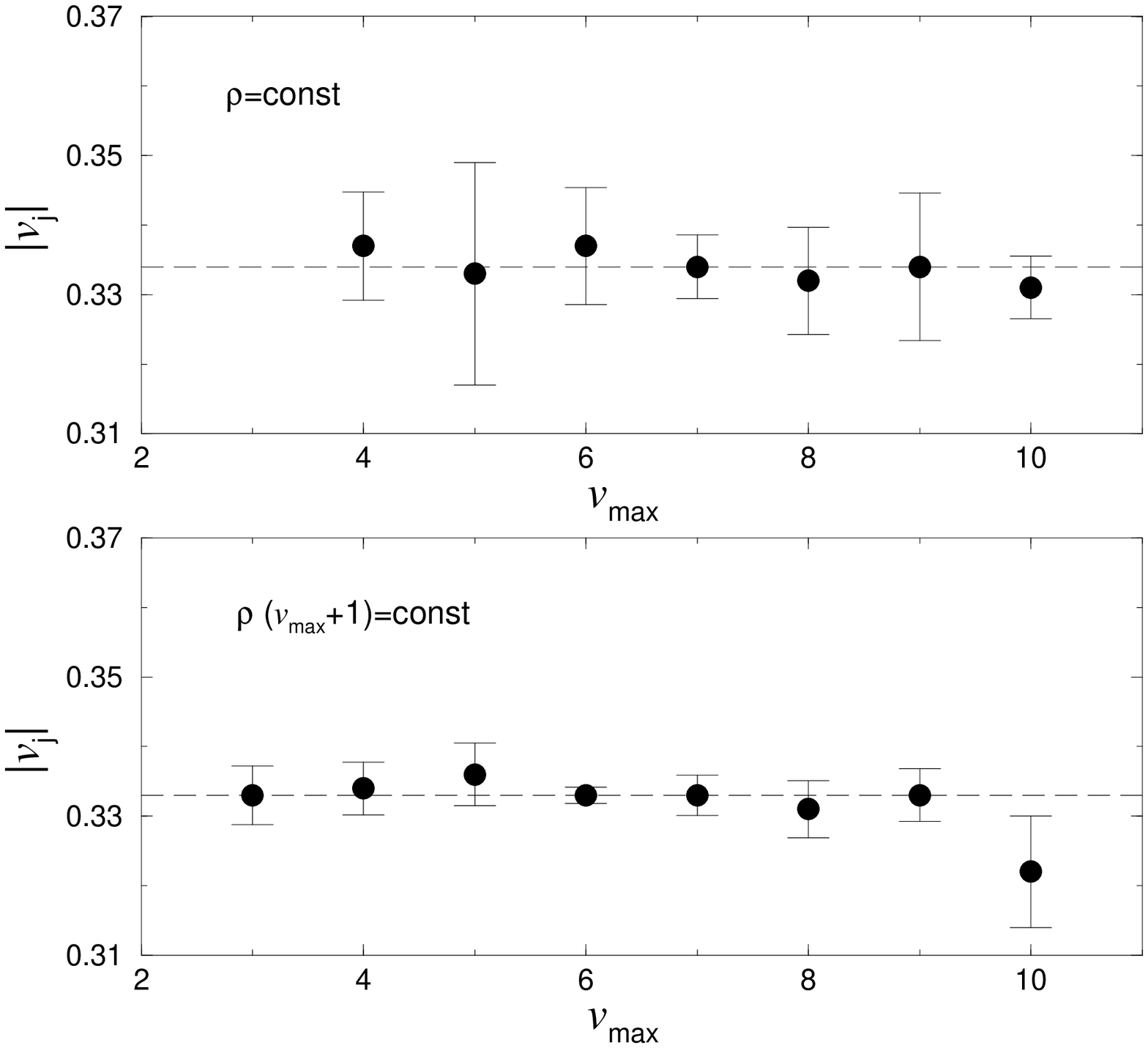} 
\end{minipage}
}}
 \caption{The characteristic velocities of the free flow 
	  phase $v_{\rm f}$ (left) and jammed phase $v_{\rm j}$
	  (right) for various values of the parameters $P$, $v_{\rm max}$, 
	  and $\rho$.
	  The dashed lines in the left figure correspond to the
	  average velocity $v_{\rm max}-P$ of a single particle.
	  The values of $v_{\rm j}$ are obtained from simulations
	  with $P=0.5$ and various values of $\rho$ and $v_{\rm max}$.}
 \label{fig:velocities} 
\end{figure}

\begin{figure}[t]
\begin{center}
 \epsfxsize=6.0cm
 \epsfysize=6.0cm
% \epsffile{/nfs/nujunkum/sven/traffic/ps_files/v_jam_03.eps} 
 \epsffile{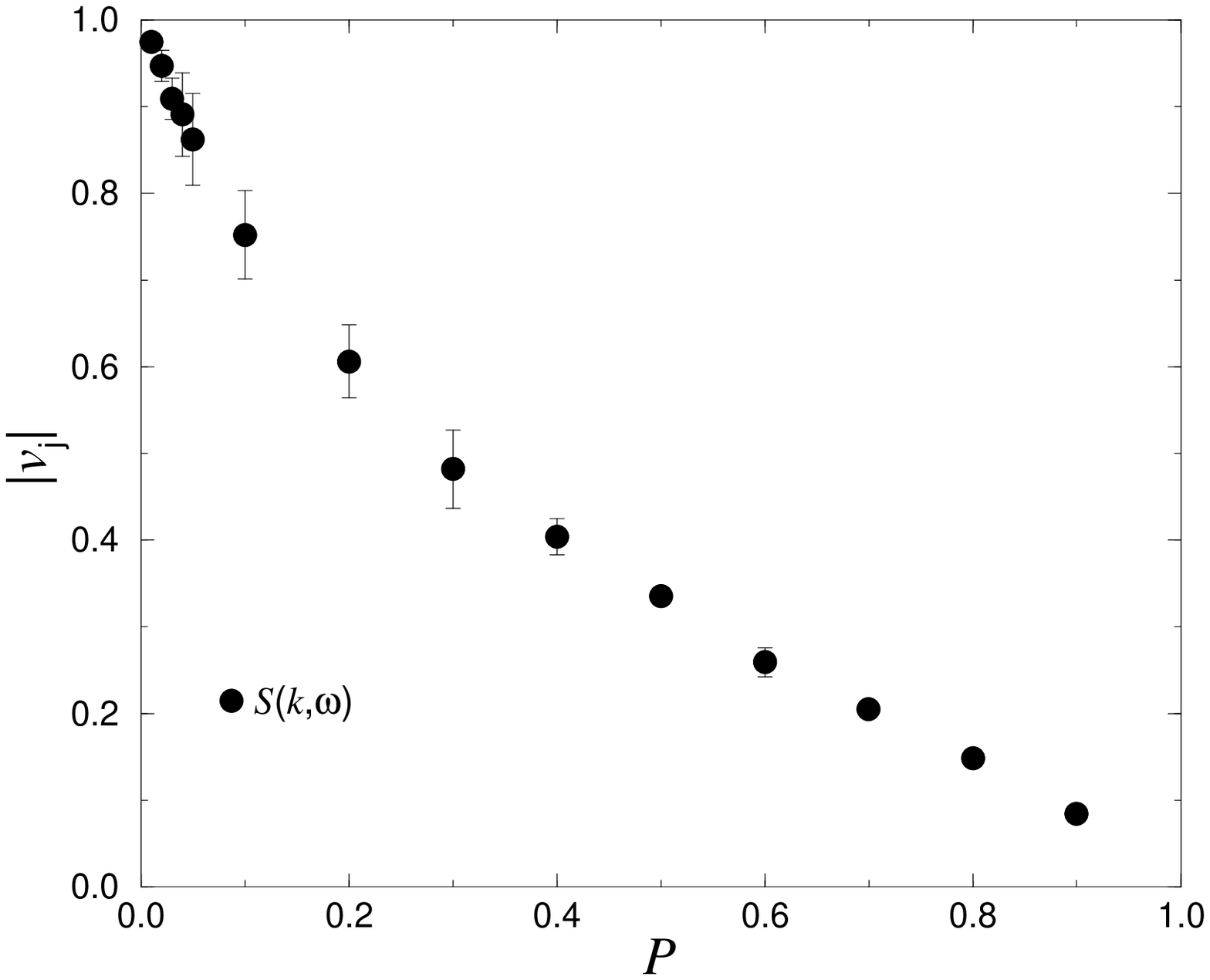} 
 \end{center}
 \vspace{-0.8cm}
 \caption{The jam velocity $v_{\rm j}$ as a function
	  of the noise parameter $P$.}
 \label{fig:jam_velocity}
\end{figure}

A linear regression of the modes of Fig.~\ref{fig:dyn_struc_factor_max}
yields the velocities of both phases.
In Fig.~\ref{fig:velocities}
we plot the velocities of the free flow and jammed phase
for different values of $P$, $v_{\rm max}$, and $\rho$.
Independent of the value of the global density, the free flow phase
is characterized by the velocity $v_{\rm f}=v_{\rm max}-P$,
i.e.~$v_{\rm f}$ equals the average velocity of the particles 
in the limit $\rho \to 0$ and the cars can be considered 
as independent particles.
This agrees with investigations of the 
local density distribution and the steady
state structure factor ($\omega = 0$)~\cite{LUEB_6}.
The plotted values of the jam velocity $v_{\rm j}$ 
are obtained from simulations with $P=0.5$ and various
values of $\rho$ and $v_{\rm max}$.
The velocity of the jams neither depends on the 
global density nor on the maximum velocity.
The jam velocity is a function of the noise parameter
only.
Using the structure of $S(k,\omega)$ we are able
to determine this $P$ dependence of $v_{\rm j}$ 
for the first time and the results are plotted
in Fig.~\ref{fig:jam_velocity}.

\section{Conclusions}

We studied numerically the Nagel-Schreckenberg traffic flow model. 
The investigation of the dynamical structure factor allowed us to examine 
the transition of the system from a free flow regime to a jammed regime. 
Above the transition the dynamical structure factor exhibits two 
modes corresponding to the coexisting free flow and jammed phase. 
Due to the sign of their characteristic velocities $v_{\rm f}$ and 
$v_{\rm j}$,
both phases can clearly be distinguished. 
We investigated the dependence of $v_{\rm f}$ and 
$v_{\rm j}$ on the system parameters 
systematically and found that the jammed velocity depends on 
the noise parameter only. 
The characteristic velocity of the free flow phase turned out to be 
equal to the velocity of free flowing cars in the low density limit 
($\rho \to 0$), i.e. cars of the free flow phase behave as independent
particles.

\end{document}